\documentclass[sigconf,%
review=false,
anonymous=false,
balance=true%
]{acmart}
\renewcommand\footnotetextcopyrightpermission[1]{} 

\usepackage{booktabs}
\setcopyright{none}

\newcommand{\X}{\textit{``X'' }}
\newcommand{\Y}{\textit{``Y'' }}
\newcommand{\A}{\textit{``A'' }}
\newcommand{\PM}{\textsc{PatentMatch}}

\begin{document}

\title{\PM: A Dataset for Matching Patent Claims \& Prior Art}

  \author{Julian Risch}
  \email{julian.risch@hpi.de}
  \affiliation{%
    \institution{Hasso Plattner Institute,}
    \city{University of Potsdam, Germany}
  }
  \author{Nicolas Alder}
  \email{nicolas.alder@student.hpi.de}
  \affiliation{%
    \institution{Hasso Plattner Institute,}
    \city{University of Potsdam, Germany}
  }

  \author{Christoph Hewel}
  \email{c.hewel@bettenpat.com}
  \affiliation{%
    \institution{BETTEN \& RESCH Patent- und Rechtsanw{\"a}lte PartGmbB,}
    \city{Munich, Germany}
  }

  \author{Ralf Krestel}
  \email{ralf.krestel@hpi.de}
  \affiliation{%
    \institution{Hasso Plattner Institute,}
    \city{University of Potsdam, Germany}
  }

\renewcommand{\shortauthors}{Risch et al.}

\begin{abstract}
  Patent examiners need to solve a complex information retrieval task when they assess the novelty and inventive step of claims made in a patent application.
  Given a claim, they search for \emph{prior art}, which comprises all relevant publicly available information.
  This time-consuming task requires a deep understanding of the respective technical domain and the patent-domain-specific language.
  For these reasons, we address the computer-assisted search for prior art by creating a training dataset for supervised machine learning called \PM.
  It contains pairs of claims from patent applications and semantically corresponding text passages of different degrees from cited patent documents. 
  Each pair has been labeled by technically-skilled patent examiners from the European Patent Office.
  Accordingly, the label indicates the degree of semantic correspondence (matching), i.e., whether the text passage is prejudicial to the novelty of the claimed invention or not. 
  Preliminary experiments using a baseline system show that \PM\ can indeed be used for training a binary text pair classifier on this challenging information retrieval task.
  The dataset is available online: \url{https://hpi.de/naumann/s/patentmatch}.
\end{abstract}



\maketitle
\pagestyle{plain}

\section{Passage Retrieval from Prior Art}
Language understanding is a very difficult task.
Even more so when considering technical, patent-domain-specific documents.
Modern deep learning approaches come close in grasping the semantic meaning of simple texts, but require a huge amount of training data.
We provide a large annotated dataset of patent claims and corresponding prior art, which not only can be used to train machine learning algorithms to recommend suitable passages to human experts, but also illustrates how experts solve this very complex IR-problem.

In general, a patent entitles the patent owner to exclude others from making, using, or selling an invention.
For this purpose, the patent comprises so-called patent claims (usually at the end of a technical description of the invention).
These claims legally specify the scope of protection of the invention.
In order to obtain a patent, it is required that the invention as defined in the claims is new and inventive over prior art~\cite{michel2001patent}.
A patent application therefore has to be filed at a patent office where it is examined on \textit{novelty} and \textit{inventive step} by a technically skilled examiner.
In case a patent is granted, said patent is published again as a separate patent document.
For this reason, there exists a huge corpus of publicly available patent documents, i.e., published patent applications and patents.

As a further consequence of this huge patent literature corpus, the examiners usually focus their prior art search on relevant patent documents.
Accordingly, they try to retrieve at least one older patent document that discloses the complete invention as defined in the claims, in particular in independent claim 1.
In other words, such a novelty-destroying document must comprise passages that semantically match with the definition of claim 1 of the examined patent application.
Said novelty-destroying document is manually marked by an expert as \X document in the search report issued by the patent office~\cite{LOVENIERS2018}.
Any retrieved document that does not disclose the complete invention defined in claim 1 but at least renders it obvious, is marked as \Y document in the search report.
Further found documents that form technological background but are not relevant to the novelty or inventive step of claim 1, are marked as \A documents. 
As a consequence, only one retrieved \X document or \Y document is enough to refuse claim 1 and hence the patent application.
Due to this circumstance, the search task is rather focused on precision than on recall.
Usually, a search report issued for an examined patent application only comprises a few (e.g., 5) cited patent documents, wherein (as far as possible) at least one document is novelty destroying (marked as \X document). 

\begin{figure*}[!htb]
\centering
{%
\setlength{\fboxsep}{0pt}%
\setlength{\fboxrule}{1pt}%
\fbox{
\includegraphics[width=0.85\textwidth]{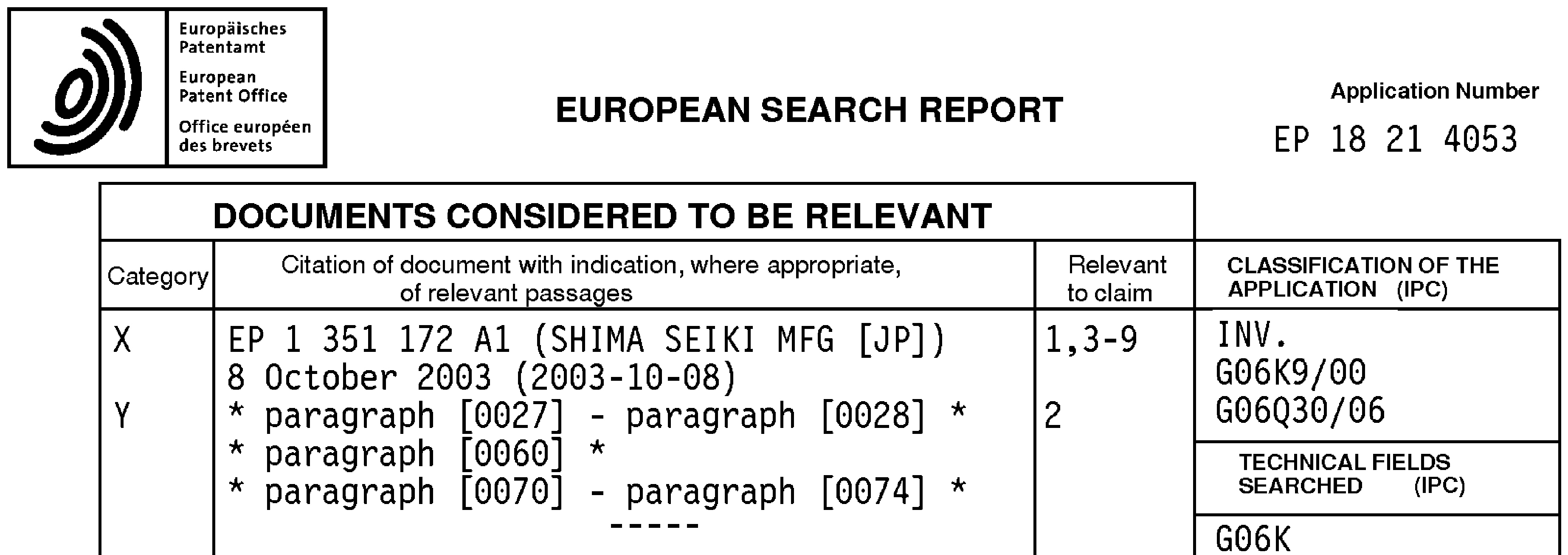}}%
}%
\caption{In this excerpt from a search report, a patent examiner cites paragraph numbers of the published patent document \textit{EP1351172A1} for assessing the novelty of claim 1 and 3-9 of application \textit{EP18214053}.}
\label{fig:search-report}
\end{figure*}

Advantageously, a search report issued by the European Patent Office (EPO) not only cites patent documents deemed relevant by an expert but also indicates for each cited document which paragraphs within the document are found to be relevant for the examined claims.
Figure~\ref{fig:search-report} exemplifies such a search report.
The EPO search report annotates each claim of the examined patent application with specific text passages (i.e., paragraphs) of a cited document.
The EPO calls this rich-format citation.
Given the application with the filing number \textit{EP18214053}, a patent officer cited prior art with the publication number \textit{EP1351172A1}.
For example, paragraph 27-28, 60 and 70-74 are relevant passages for assessing the novelty of claims 1 and 3 to 9.
The search report also lists which search terms were used. 
In this case, it is the IPC subclass \textit{G06K}.

\section{Related Work}
Finding relevant prior art is even for well-trained experts a hard and cumbersome task~\cite{jeffery2002preserving}.
Due to the large volume of literature to be considered as well as the required domain knowledge, patent officers rely on modern information systems to support them with their task~\cite{martin2018}.
Nevertheless, the outcome of a prior art search, either to check for patentability or validity of a patent, remains imperfect and biased based on the patent examiner and her search strategy~\cite{lei2017}.
In addition, different patent offices can reach different conclusions for the same search~\cite{michel2001patent}.
With this paper we hope to open the door to qualitatively and systematically analyse the search practice particularly at European Patent Office.

Traditionally, related work at the intersection of information retrieval and patent analysis aims to support the experts by automatically identifying technical terms in patent documents~\cite{judea2014unsupervised} or keywords that relate to the novelty of claims in applications~\cite{suzuki2016extraction}.
A challenge that all natural language processing applications in the patent domain have is to cope with the legal jargon and specialized terminology, which led to the use of patent-domain-specific word embeddings in deep learning approaches~\cite{risch2019domainspecific,abdelgawad2019optimizing}.
Further, patent classification is the most prominent task for the application of natural language processing in this domain, with supervised deep learning approaches outperforming all other methods~\cite{Li2018,risch2019domainspecific}.
Large amounts of labeled training data are available for this task because every published patent document and application is classified according to standardized, hierarchical classification schemes.

Prior art search is a document retrieval task where the goal is to find related work for a given patent document or application.
Formulating the corresponding search query is a research challenge typically addressed with keyword extraction~\cite{golestan2015term,xue2009transforming,tseng2008study}.
Further, there is research on tools to support expert users in defining search queries~\cite{russellrose2019visual} or non-expert users in exploring the search space step by step~\cite{kulahcioglu2017incorporating}.
The task that we focus on in this paper is patent passage retrieval.
Given a query passage, e.g., a claim, the task is to find relevant passages in a corpus of text documents to, e.g., decide on the novelty of the claim.
In the CLEF-IP series of shared tasks, there was a \emph{claims to passage task} in 2012~\cite{gobeill2012bitem,piroi2012retrieval}.
The shared task dataset contains 2.3 million documents and 2700 relevance judgements of passages for training, which were manually extracted from search reports.
The passages are contained in \X documents and \Y documents referenced by patent examiners in the search reports.
Similar passage retrieval tasks can be found in other domains as well, e.g., passage retrieval for question answering within Wikipedia~\cite{cohen2018wikipassageqa}.

Research in the patent domain is limited for three reasons: patent-domain-specific knowledge is necessary to understand (1) different types of documents (patent applications, granted patents, search reports), (2) different classification schemes (IPC, CPC, USPC) and (3) the steps of the patenting process (filing, examination, publication, granting, opposition).

In this paper, we present \PM, a dataset of claims from patent applications matched with paragraphs from prior art, e.g., published patent documents.
Professional patent examiners labeled the claims with references to paragraphs that are prejudicial to the novelty of the claim (\X documents, positive samples) or that are not prejudicial but represent merely technical background (\A documents, negative samples).
We collected these labels from search reports created by patent examiners, resolved the claims and paragraphs referenced therein, and extracted the corresponding text passages from the patent documents. 
This procedure resulted in a dataset of six million examined claims and semantically corresponding (matching) text passages that are prejudicial or not prejudicial to the novelty of the claims.
The remainder of this paper is structured as follows: 
Section~\ref{sec:dataset} describes the data collection and processing steps in detail and provides dataset examples and statistics.
Section~\ref{sec:tasks} outlines research tasks that could benefit from the dataset and presents a preliminary experiment for one of these tasks.
Finally, Section~\ref{sec:conclusions} concludes with a discussion of the potential impact of the presented dataset.

\section{\PM~Dataset}
\label{sec:dataset}
The basis of our dataset is the \emph{EP full-text data for text analytics} by the EPO.\footnote{\url{https://www.epo.org/searching-for-patents/data/bulk-data-sets/text-analytics}} 
It contains the XML-formatted full-texts and publication meta-data of all filed patent applications and published patent documents processed by the EPO since 1978. 
From 2012 onwards, the search reports for all patent applications are also included. 
In these reports, patent examiners cite paragraphs from prior art documents if these paragraphs are relevant for judging the novelty and inventive step of an application claim.
Although there are no search reports available for applications filed before 2012, we do not discard these older applications because their corresponding published patent documents are frequently referenced as prior art. 
We use all available search reports to create a dataset of claims of patent applications matched with prior art, more precisely, paragraphs of cited \X documents and \A documents.

Our data processing pipeline uses Elasticsearch for storing and searching through this large corpus of about 210GB of text data.
As a first data preparation step, an XML parser extracts the full text and meta-data from the raw, multi-nested XML files.
Further, for each citation within a search report, it extracts claim number, patent application ID, date, paragraph number, and the type of the references, i.e., \X document or \A document.

Since the search reports were written in a rather systematic, but still unstructured and non-consistent way, a second parsing step standardizes the data format of paragraph references. 
References like ``[paragraph 23]-[paragraph 28]'' or ``0023 - 28'' are converted to complete enumerations of paragraph numbers ``[23,24,25,26,27,28]''. 
Furthermore, references by patent examiners comprise not only text paragraphs but also figures, figure captions, or the whole document.
In our standardization process, all references that do not resolve to text paragraphs are discarded.

\begin{figure}[ht]
\centering
\includegraphics[width=\linewidth]{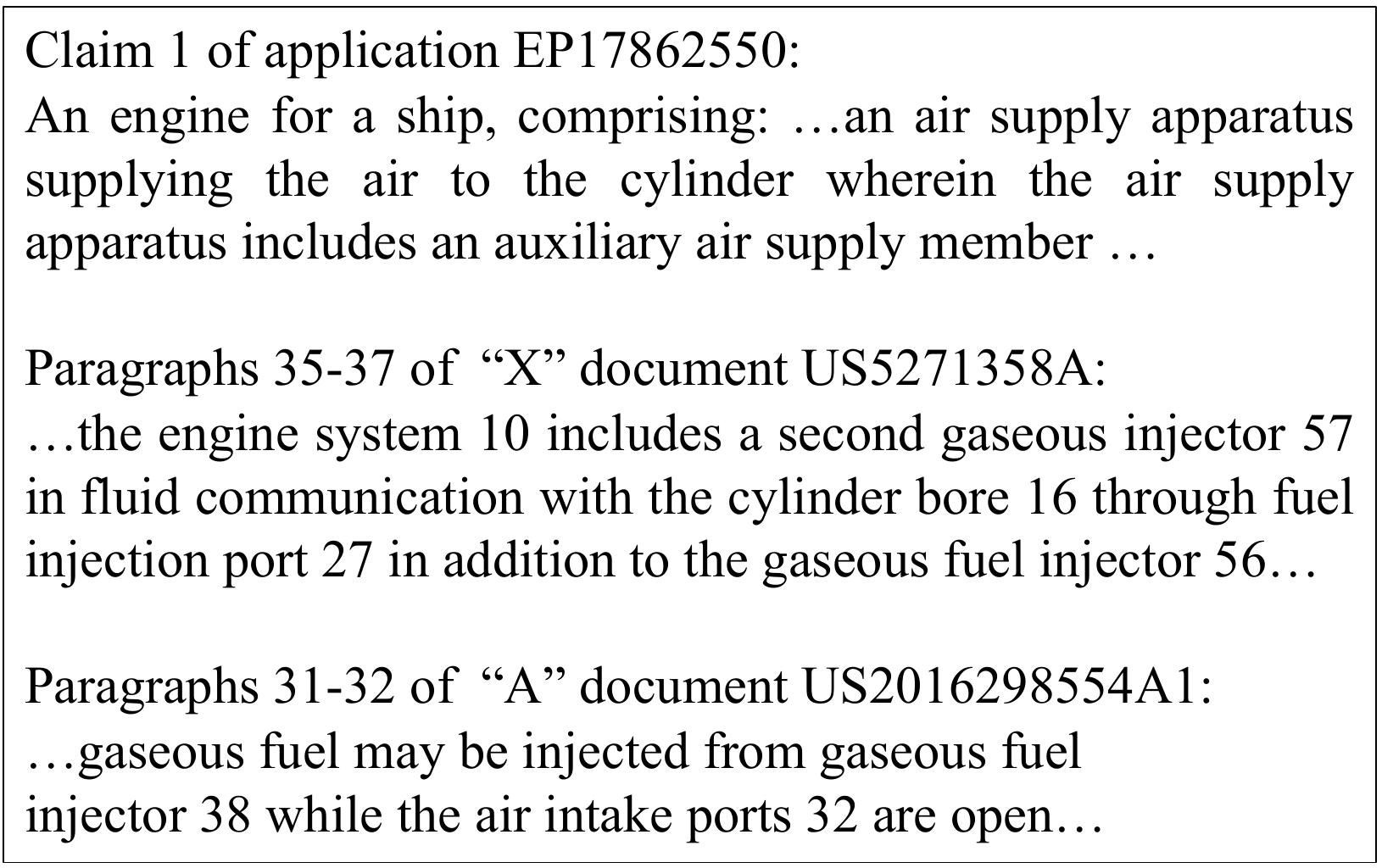}
\caption{An excerpt from a search report showing a claim and cited paragraphs. The \X document (positive sample) is novelty-destroying for the claim while the \A document (negative sample) is not novelty-destroying and merely constitutes technical background.}
\label{fig:search}
\end{figure}

In the final step, we use the index of our Elasticsearch document database to resolve the referenced paragraph numbers (together with the corresponding document identifiers) to the paragraph texts.
Similarly, we resolve the claim texts corresponding to the claim numbers. 
Thereby, we obtain a dataset that consists of a total of 6,259,703 samples, where each sample contains a claim text, a referenced paragraph text, and a label indicating one of the two types of reference: \X document (positive sample) or \A document (negative sample).
Table~\ref{tab:dataset} lists statistics of the full dataset and Figure~\ref{fig:search} exemplifies a claim text and cited paragraph texts of positive and negative samples.
\begin{table}[tb]
\caption{Dataset statistics: Each sample is a pair of an application's claim and paragraph cited from either an \X document (positive sample) or \A document (negative sample).}
\label{tab:dataset}
\centering
\begin{tabular}{lr}
\toprule
Samples & 6,259,703\\
\X document citations & 3,492,987\\
\A document citations & 2,766,716\\
\midrule
Distinct patent applications & 31,238\\
Distinct cited documents & 33,195\\
Distinct claim texts & 297,147\\
Distinct cited paragraphs & 520,376\\
\midrule
Median claim length (chars) & 274 \\
Median paragraph length (chars) & 476 \\
\bottomrule
\end{tabular}
\end{table}

We also provide two variations of the data for simplified usage in machine learning scenarios. 
The first variation balances the label distributions by downsampling the majority class. 
For each sample with a claim text and a referenced paragraph labeled \textit{``X''}, there is also a sample with the same claim text with a different referenced paragraph labeled \A and vice versa.
This balanced training set consists of 347,880 samples.
In this version of the dataset, different claim texts can have different numbers of references.
The number of \X and \A labels is only balanced for each claim text itself.

The second variation balances not only the label distribution but also the distribution of claim texts. 
Further downsampling ensures that there is exactly one sample with label \X and one sample with label \A for each claim text.
As a result, every claim in the dataset occurs in exactly two samples. 
This restriction reduces the dataset to 25,340 samples. 

The \PM~dataset is published online with example code that shows how to use it for supervised machine learning, and a description of the data collection and preparation process.\footnote{\url{https://hpi.de/naumann/s/patentmatch}}
As the underlying raw data has been released by the EPO under Creative Commons Attribution 4.0 International Public License, we also release our dataset under the same license.\footnote{\url{https://creativecommons.org/licenses/by/4.0/}}
To foster comparable evaluation settings in future work, we separated it into a training set (80\%) and a test set (20\%) with a time-wise split based on the application filing date:
All applications contained in the training set have an earlier filing date than all applications contained in the test set.

\section{Preliminary Experiments}
\label{sec:tasks}
Modern information retrieval systems do not solely rely on matching keywords from queries with documents.
Especially for complex information needs, semantic knowledge needs to be incorporated~\cite{fernandez2011semantically}.
With the rise of deep learning models, as well as word and document embeddings, improvements in grasping the semantic meaning of queries and documents have been made~\cite{cohen18hybrid}.
A number of related tasks aim at finding semantically related information, making use of advanced semantic representations~\cite{ganguly2015word} and intelligent retrieval models~\cite{palangi2016deep}.
Passage retrieval~\cite{kaszkiel97passage}, document clustering~\cite{huang2008similarity}, and question answering~\cite{yang2017efficiently} all rely on identifying semantically related information.

Addressing a first exemplary task, we conducted preliminary experiments on text pair classification with Bidirectional Encoder Representations from Transformers (BERT)~\cite{devlin2018bert} as a baseline system.
The text pair classification uses the same neural network architecture as the next sentence prediction task:
Given a pair of sentences, the next sentence prediction task is to predict if the second sentence is a likely continuation of the first sentence.
In our text pair classification scenario, given a claim text and a cited paragraph text, the task is to decide whether the paragraph corresponds to an \X document (positive sample) or an \A document (negative sample).
To make this decision, the model needs to assess the novelty of the claim in comparison to the paragraph.
To this end, it transforms the input text to sub-word tokens and transforms them to their embedding representations.
These representation pass through 12 layers of bidirectional Transformers~\cite{vaswani2017attention} and the final hidden state of the special token \texttt{[CLS]} encodes the output class label.
Our implementation uses the FARM framework and the pre-trained bert-base-uncased model.\footnote{\url{https://github.com/deepset-ai/FARM}\textsc{,} \url{https://huggingface.co/bert-base-uncased}}

The test set accuracy on the balanced variation of the data is 54\%. 
On the second variation of the data, which contains exactly one \X document citation and one \A document citation per claim, the accuracy on the test set is 52\%.
For both variations, the accuracy improvements per training epoch are small and the validation loss stops to decrease after training for 6 epochs.
It is not to our surprise that the task poses a difficult challenge and that a fine-tuned BERT model is only slightly better than random guessing.
The complex linguistic patterns, the legal jargon, and the patent-domain-specific language make it sheer impossible for laymen to manually solve this task and therefore an interesting research challenge for future work.

\section{Impact \& Conclusions}
\label{sec:conclusions}
With this paper, we not only introduce an extensive dataset that can be used to train and test systems for the aforementioned tasks, but also provide training data for patent passage retrieval~\cite{piroi2012retrieval}: a very challenging search task mostly conducted by highly-trained patent-domain experts.
The need to at least partially automate this task arises from the growing number of patent applications worldwide.

And with deep learning methods requiring large training sets, we hope to foster research in the patent analysis domain by providing such a dataset.
We presented a novel dataset that comprises pairs of semantically similar texts in the patent domain.
More precisely, the dataset contains claims from patent applications and paragraphs from prior art.
It was created based on search reports by patent officers at the EPO.
The simple structure of the dataset reduces the amount of patent-domain knowledge required for analyzing the data or using it for supervised machine learning.
With the release of the dataset, we thus hope to foster research on the (semi-)automation of passage retrieval tasks and on user interfaces that support experts in searching through prior art and creating search reports.

Further, we hope to spark research in analysing how patent experts search for relevant patents and, maybe more interesting, which relevant patents they miss and for what reason.
By providing the matched claims and paragraphs, the search process of patent officers can be analyzed and search results compared.
For future work, our learned model could be used to adapt the experts' keyword queries for higher recall and to understand the relationship between results from manually curated queries and (relevant) results from deep learning models.

\begin{acks}
We would like to thank Sonia Kaufmann and Martin Kracker from the European Patent Office (EPO) for their support and advise.
\end{acks}

\bibliographystyle{abbrvnat}
\bibliography{acmart}

\begin{thebibliography}{27}
\providecommand{\natexlab}[1]{#1}
\providecommand{\url}[1]{\texttt{#1}}
\expandafter\ifx\csname urlstyle\endcsname\relax
  \providecommand{\doi}[1]{doi: #1}\else
  \providecommand{\doi}{doi: \begingroup \urlstyle{rm}\Url}\fi

\bibitem[Abdelgawad et~al.(2019)Abdelgawad, Kluegl, Genc, Falkner, and
  Hutter]{abdelgawad2019optimizing}
L.~Abdelgawad, P.~Kluegl, E.~Genc, S.~Falkner, and F.~Hutter.
\newblock Optimizing neural networks for patent classification.
\newblock In \emph{Joint European Conference on Machine Learning and Knowledge
  Discovery in Databases (ECML PKDD)}, pages 688--703, 2019.

\bibitem[Cohen and Croft(2018)]{cohen18hybrid}
D.~Cohen and W.~B. Croft.
\newblock A hybrid embedding approach to noisy answer passage retrieval.
\newblock In \emph{Advances in Information Retrieval}, pages 127--140, 2018.

\bibitem[Cohen et~al.(2018)Cohen, Yang, and Croft]{cohen2018wikipassageqa}
D.~Cohen, L.~Yang, and W.~B. Croft.
\newblock {WikiPassageQA}: A benchmark collection for research on non-factoid
  answer passage retrieval.
\newblock In \emph{Proceedings of the International Conference on Research and
  Development in Information Retrieval (SIGIR)}, page 1165–1168, 2018.

\bibitem[Devlin et~al.(2018)Devlin, Chang, Lee, and Toutanova]{devlin2018bert}
J.~Devlin, M.-W. Chang, K.~Lee, and K.~Toutanova.
\newblock Bert: Pre-training of deep bidirectional transformers for language
  understanding.
\newblock \emph{arXiv preprint arXiv:1810.04805}, pages 1--16, 2018.

\bibitem[Fern{\'a}ndez et~al.(2011)Fern{\'a}ndez, Cantador, L{\'o}pez, Vallet,
  Castells, and Motta]{fernandez2011semantically}
M.~Fern{\'a}ndez, I.~Cantador, V.~L{\'o}pez, D.~Vallet, P.~Castells, and
  E.~Motta.
\newblock Semantically enhanced information retrieval: An ontology-based
  approach.
\newblock \emph{Journal of Web Semantics}, 9\penalty0 (4):\penalty0 434--452,
  2011.

\bibitem[Ganguly et~al.(2015)Ganguly, Roy, Mitra, and Jones]{ganguly2015word}
D.~Ganguly, D.~Roy, M.~Mitra, and G.~J. Jones.
\newblock Word embedding based generalized language model for information
  retrieval.
\newblock In \emph{Proceedings of the International Conference on Research and
  Development in Information Retrieval (SIGIR)}, pages 795--798, 2015.

\bibitem[Gobeill and Ruch(2012)]{gobeill2012bitem}
J.~Gobeill and P.~Ruch.
\newblock Bitem site report for the claims to passage task in {CLEF-IP} 2012.
\newblock In \emph{Proceedings of the {CLEF-IP} Workshop}, 2012.

\bibitem[Golestan~Far et~al.(2015)Golestan~Far, Sanner, Bouadjenek, Ferraro,
  and Hawking]{golestan2015term}
M.~Golestan~Far, S.~Sanner, M.~R. Bouadjenek, G.~Ferraro, and D.~Hawking.
\newblock On term selection techniques for patent prior art search.
\newblock In \emph{Proceedings of the International Conference on Research and
  Development in Information Retrieval (SIGIR)}, pages 803--806, 2015.

\bibitem[Huang(2008)]{huang2008similarity}
A.~Huang.
\newblock Similarity measures for text document clustering.
\newblock In \emph{Proceedings of the New Zealand Computer Science Research
  Student Conference (NZCSRSC)}, volume~4, pages 9--56, 2008.

\bibitem[Jeffery(2002)]{jeffery2002preserving}
J.~A. Jeffery.
\newblock Preserving the presumption of patent validity: An alternative to
  outsourcing the us patent examiner's prior art search.
\newblock \emph{Cath. UL Rev.}, 52:\penalty0 761, 2002.

\bibitem[Judea et~al.(2014)Judea, Sch{\"u}tze, and
  Br{\"u}gmann]{judea2014unsupervised}
A.~Judea, H.~Sch{\"u}tze, and S.~Br{\"u}gmann.
\newblock Unsupervised training set generation for automatic acquisition of
  technical terminology in patents.
\newblock In \emph{Proceedings of the International Conference on Computational
  Linguistics (COLING)}, pages 290--300, 2014.

\bibitem[Kaszkiel and Zobel(1997)]{kaszkiel97passage}
M.~Kaszkiel and J.~Zobel.
\newblock Passage retrieval revisited.
\newblock In \emph{Proceedings of the International Conference on Research and
  Development in Information Retrieval (SIGIR)}, pages 178--185, 1997.

\bibitem[Kulahcioglu et~al.(2017)Kulahcioglu, Fradkin, and
  Palanivelu]{kulahcioglu2017incorporating}
T.~Kulahcioglu, D.~Fradkin, and S.~Palanivelu.
\newblock Incorporating task analysis in the design of a tool for a complex and
  exploratory search task.
\newblock In \emph{Proceedings of the Conference on Conference Human
  Information Interaction and Retrieval (CHIIR)}, page 373–376, 2017.

\bibitem[Lei and Wright(2017)]{lei2017}
Z.~Lei and B.~D. Wright.
\newblock Why weak patents? testing the examiner ignorance hypothesis.
\newblock \emph{Journal of Public Economics}, 148:\penalty0 43 -- 56, 2017.
\newblock ISSN 0047-2727.
\newblock \doi{https://doi.org/10.1016/j.jpubeco.2017.02.004}.
\newblock URL
  \url{http://www.sciencedirect.com/science/article/pii/S0047272717300178}.

\bibitem[Li et~al.(2018)Li, Hu, Cui, and Hu]{Li2018}
S.~Li, J.~Hu, Y.~Cui, and J.~Hu.
\newblock Deeppatent: patent classification with convolutional neural networks
  and word embedding.
\newblock \emph{Scientometrics}, 117\penalty0 (2):\penalty0 721--744, 2018.

\bibitem[Loveniers(2018)]{LOVENIERS2018}
K.~Loveniers.
\newblock How to interpret epo search reports.
\newblock \emph{World Patent Information}, 54:\penalty0 23--28, 2018.

\bibitem[Marttin and Derrien(2018)]{martin2018}
E.~Marttin and A.-C. Derrien.
\newblock How to apply examiner search strategies in espacenet. a case study.
\newblock \emph{World Patent Information}, 54:\penalty0 S33 -- S43, 2018.
\newblock ISSN 0172-2190.
\newblock \doi{https://doi.org/10.1016/j.wpi.2017.06.001}.
\newblock URL
  \url{http://www.sciencedirect.com/science/article/pii/S0172219016301089}.
\newblock Best of Search Matters.

\bibitem[Michel and Bettels(2001)]{michel2001patent}
J.~Michel and B.~Bettels.
\newblock Patent citation analysis. a closer look at the basic input data from
  patent search reports.
\newblock \emph{Scientometrics}, 51\penalty0 (1):\penalty0 185--201, 2001.

\bibitem[Palangi et~al.(2016)Palangi, Deng, Shen, Gao, He, Chen, Song, and
  Ward]{palangi2016deep}
H.~Palangi, L.~Deng, Y.~Shen, J.~Gao, X.~He, J.~Chen, X.~Song, and R.~Ward.
\newblock Deep sentence embedding using long short-term memory networks:
  Analysis and application to information retrieval.
\newblock \emph{IEEE/ACM Transactions on Audio, Speech, and Language
  Processing}, 24\penalty0 (4):\penalty0 694--707, 2016.

\bibitem[Piroi et~al.(2012)Piroi, Lupu, Hanbury, Sexton, Magdy, and
  Filippov]{piroi2012retrieval}
F.~Piroi, M.~Lupu, A.~Hanbury, A.~P. Sexton, W.~Magdy, and I.~V. Filippov.
\newblock {CLEF-IP} 2012: Retrieval experiments in the intellectual property
  domain.
\newblock In \emph{Proceedings of the {CLEF-IP} Workshop}, pages 1--16, 2012.

\bibitem[Risch and Krestel(2019)]{risch2019domainspecific}
J.~Risch and R.~Krestel.
\newblock Domain-specific word embeddings for patent classification.
\newblock \emph{Data Technologies and Applications}, 53\penalty0 (1):\penalty0
  108--122, 2019.

\bibitem[Russell-Rose et~al.(2019)Russell-Rose, Chamberlain, and
  Shokraneh]{russellrose2019visual}
T.~Russell-Rose, J.~Chamberlain, and F.~Shokraneh.
\newblock A visual approach to query formulation for systematic search.
\newblock In \emph{Proceedings of the Conference on Human Information
  Interaction and Retrieval (CHIIR)}, page 379–383, 2019.

\bibitem[Suzuki and Takatsuka(2016)]{suzuki2016extraction}
S.~Suzuki and H.~Takatsuka.
\newblock Extraction of keywords of novelties from patent claims.
\newblock In \emph{Proceedings of the International Conference on Computational
  Linguistics (COLING)}, pages 1192--1200, 2016.

\bibitem[Tseng and Wu(2008)]{tseng2008study}
Y.-H. Tseng and Y.-J. Wu.
\newblock A study of search tactics for patentability search: A case study on
  patent engineers.
\newblock In \emph{Proceedings of the Workshop on Patent Information Retrieval
  (PaIR@CIKM)}, pages 33--36, 2008.

\bibitem[Vaswani et~al.(2017)Vaswani, Shazeer, Parmar, Uszkoreit, Jones, Gomez,
  Kaiser, and Polosukhin]{vaswani2017attention}
A.~Vaswani, N.~Shazeer, N.~Parmar, J.~Uszkoreit, L.~Jones, A.~N. Gomez,
  {\L}.~Kaiser, and I.~Polosukhin.
\newblock Attention is all you need.
\newblock In \emph{Advances in Neural Information Processing Systems
  (NeurIPS)}, pages 5998--6008, 2017.

\bibitem[Xue and Croft(2009)]{xue2009transforming}
X.~Xue and W.~B. Croft.
\newblock Transforming patents into prior-art queries.
\newblock In \emph{Proceedings of the International Conference on Research and
  Development in Information Retrieval (SIGIR)}, pages 808--809, 2009.

\bibitem[Yang et~al.(2017)Yang, Zou, Wang, Yan, and Wen]{yang2017efficiently}
S.~Yang, L.~Zou, Z.~Wang, J.~Yan, and J.-R. Wen.
\newblock Efficiently answering technical questions—a knowledge graph
  approach.
\newblock In \emph{Proceedings of the Conference on Artificial Intelligence
  (AAAI)}, pages 3111--3118, 2017.

\end{thebibliography}

\end{document}